\documentclass[11pt]{amsart}

\usepackage{fullpage}

\newtheorem{theorem}{Theorem}[section]

\newtheorem{corollary}[theorem]{Corollary}

\newtheorem{lemma}[theorem]{Lemma}

\newtheorem{proposition}[theorem]{Proposition}

\newtheorem{remark}[theorem]{Remark}

\newtheorem{definition}[theorem]{Definition}

\def\endproof{\qed \medskip}
\def\blacksquare{\hbox to .60em {\vrule width .60em height .60em}}

\begin{document}

\title[]{On the Uniqueness and Global Dynamics of AdS Spacetimes}

\author[]{Michael T. Anderson}

\thanks{Partially supported by NSF Grant DMS 0305865}

\abstract
{We study global aspects of complete, non-singular asymptotically locally 
AdS spacetimes solving the vacuum Einstein equations whose conformal infinity 
is an arbitrary globally stationary spacetime. It is proved that any such 
solution which is asymptotically stationary to the past and future is 
itself globally stationary. 

  This gives certain rigidity or uniqueness results for exact AdS and 
related spacetimes. }
\endabstract

\maketitle

\setcounter{section}{0}

\section{Introduction}
\setcounter{equation}{0}

 Consider geodesically complete, asymptotically simple solutions of the 
vacuum Einstein equations with negative cosmological constant $\Lambda < 0$ 
in $(n+1)$ dimensions. Up to rescaling, these are given by complete, 
(non-singular), metrics $g$, defined on manifolds of the form 
$M^{n+1} = {\mathbb R} \times \Sigma$, and satisfying the Einstein equations
\begin{equation} \label{e1.1}
Ric_{g} = -ng. 
\end{equation}
The metric $g$ has a conformal completion, at least $C^{2}$, with 
conformal boundary $({\mathcal I}, [\gamma])$, where $\gamma$ is a 
complete Lorentz metric on ${\mathcal I}$. Topologically, conformal 
infinity ${\mathcal I}$ is of the form ${\mathbb R} \times \partial\Sigma$. 

 The canonical example is the (exact) anti-de Sitter spacetime 
$g_{AdS}$, which may be represented globally in static form as 
\begin{equation} \label{e1.2}
g_{AdS} = -\cosh^{2}r\, dt^{2} + dr^{2} + \sinh^{2}r\, g_{S^{n-1}(1)}, 
\end{equation}
where $g_{S^{n-1}(1)}$ is the round metric of radius 1 on the sphere 
$S^{n-1}$. Here $M = {\mathbb R} \times {\mathbb R}^{n}$, with conformal infinity 
${\mathcal I}  = {\mathbb R}\times S^{n-1}$, with boundary metric $\gamma_{0} = 
-dt^{2} + g_{S^{n-1}(1)}$ the Einstein static cylinder. Asymptotically 
simple spacetimes approximate the metric $g_{AdS}$ locally on approach 
to ${\mathcal I}$, and so are often also called asymptotically locally AdS 
spacetimes. 

 It is generally believed that anti-de Sitter spacetime should have 
an infinite dimensional space of dynamical perturbations, i.e. 
time-dependent, complete vacuum solutions (1.1), which have the same 
conformal infinity $({\mathcal I}, \gamma_{0})$ as exact AdS and which 
are globally close to $g_{AdS}$. In other words, $g_{AdS}$ is dynamically 
stable. This is certainly the case at the linearized level; one may 
globally solve the linearized Einstein equations at $g_{AdS}$, with 
zero boundary data on ${\mathcal I}$ and arbitrary smooth Cauchy data on 
$\Sigma$. These linearized solutions, (or normalizable modes), remain 
uniformly bounded in time, cf. [1] for a detailed treatment. 

 Such dynamical or global stability results are well-known in the 
context of spacetimes with zero cosmological constant, $\Lambda  = 0$. 
Thus, the celebrated work of Christodoulou-Klainerman [2] shows that 
there exist global non-singular perturbations of Minkowski spacetime, 
(in 3+1 dimensions), which tend to the flat Minkowski spacetime as 
$t \rightarrow  \pm\infty$; see also [3] and the more recent work 
[4], as well as [5], which gives the existence of non-singular 
asymptotically simple global perturbations. There are also analogues 
of such stability results in the context of cosmological spacetimes, 
(in the expanding direction), where $\Sigma$ is compact without 
boundary, cf.~[6], [7]. 

 Similarly, in the context where $\Lambda > 0$, Friedrich [8] has 
proved that exact de Sitter spacetime is globally stable in a natural 
sense in $3+1$ dimensions; the same result holds in fact at least in 
all even dimensions, cf.~[9]. 

 In this paper, we discuss some aspects of the global dynamics of complete, 
asymptotically simple solutions of the Einstein equations when $\Lambda < 0$. 
To describe the main results, let $\gamma$ be a fixed but arbitrary globally 
stationary metric on ${\mathcal I} \simeq {\mathbb R}\times \partial \Sigma$,  
with $\partial \Sigma$ compact. Let ${\mathcal E} = {\mathcal E}(\gamma)$ be 
the space of all geodesically complete, asymptotically simple solutions 
of the Einstein equations (1.1) with conformal infinity $({\mathcal I}, 
\gamma)$; (this definition will be made more precise in \S 2). Choose a 
fixed diffeomorphism $M \simeq {\mathbb R}\times \Sigma$, (i.e. spacetime 
decomposition), and let ${\mathcal C}$ denote the corresponding space of 
solutions of the constraint equations on $(\Sigma, g, K)$ given by solutions 
in ${\mathcal E}$. In particular, a metric $g \in E$ gives rise to a curve 
$(g_{t}, K_{t})$, $t \in (-\infty, \infty)$ in ${\mathcal C}$.

\begin{theorem} \label{t1.1} 
Suppose $g \in {\mathcal E}$ is weakly asymptotically stationary in the sense 
that there exist times $t_{i} \rightarrow \infty$ and $t_{i} \rightarrow -\infty$ 
such that $(g_{t_{i}}, K_{t_{i}})$ converges to a globally stationary solution of 
the vacuum constraint equations, as $i \rightarrow \infty$. Suppose further that, 
modulo infinitesimal diffeomorphisms, solutions of the linearized Einstein 
equations at $g$ satisfy the unique continuation property at ${\mathcal I}$. 

 Then $(M^{n+1}, g)$ is globally stationary. 

\end{theorem}

  As explained in detail in \S 2, the unique continuation property at ${\mathcal I}$ 
means that solutions of the linearized Einstein equations are uniquely determined, up 
to infinitesimal diffeomorphisms, by their Cauchy data, ({\em not} just the boundary data), 
at ${\mathcal I}$. Equivalently, they are uniquely determined, up to infinitesimal 
diffeomorphism, by their formal series expansion at ${\mathcal I}$. This is 
automatically the case for solutions of the linearized equations which are 
analytic, (in the polyhomogeneous sense), at ${\mathcal I}$. The unique 
continuation property was proved to hold in general for Riemannian-Einstein 
metrics in [10], and we conjecture that it also always holds for Lorentzian-Einstein 
metrics. There certainly appears to be no physical reasons to doubt the validity of 
this property. 

\medskip

  The time evolution of metrics $g \in E$ gives rise to a flow, (i.e.~dynamical system), 
on the constraint space ${\mathcal C}$. Stationary solutions in ${\mathcal E}$ then 
correspond to fixed points of this flow, (for a suitable choice of spacetime foliation 
of $M$). Roughly speaking, Theorem 1.1 then states that there are no orbits of the flow 
on ${\mathcal C}$ which are weakly asymptotic to a fixed point at both ends, except 
fixed orbits. 

  It is possible that $({\mathcal I}, \gamma)$ has more than one timelike Killing field, 
(modulo constants). This will occur when there are non-trivial spatial Killing fields 
on $({\mathcal I}, \gamma)$. Theorem 1.1 applies to each Killing field and leads easily 
to the following corollary. 

\begin{corollary} \label{c1.2}
Under the assumptions in Theorem 1.1, suppose $({\mathcal I}, \gamma)$ is the 
Einstein static cylinder and that $(M^{n+1}, g)$ is weakly asymptotic to 
the exact AdS spacetime to the infinite past and infinite future. Then $(M^{n+1}, g)$ 
is globally isometric to the exact AdS spacetime. 
\end{corollary}

  Similar rigidity or uniqueness results hold for other spacetimes which, for example, 
have sufficient symmetry at conformal infinity $({\mathcal I}, \gamma)$. Thus, the 
Horowitz-Myers AdS soliton [11] is unique in the sense above, cf.~ \S3. The same 
applies to the AdS soliton metric of Copsey-Horowitz [12]. Previously, the uniqueness 
of the Horowitz-Myers soliton metric was only known within the much more restrictive 
class of static solutions of the Einstein equations, cf.~[13], [14]. 

  A more general result on the extension of isometries from the boundary to the bulk 
is proved in [10] for Riemannian (or Euclidean) Einstein metrics (1.1), by completely 
different methods. 

\medskip

  The proofs of the results above are conceptually very simple. They follow from 
basic conservation properties of the holographic stress-energy tensor and 
holographic mass arising in the AdS/CFT correspondence, [15], [16], [17], [18], 
[19], together with a basic identity discussed in \S 3, (cf. (3.1)). For simplicity, 
we have restricted the analysis to vacuum solutions of the Einstein equations 
(1.1) with negative cosmological constant. Of course the results above also 
hold in the presence of matter terms which arise from a vacuum solution in 
higher dimensions via Kaluza-Klein reduction. However, in general the presence 
of matter terms changes the conservation properties of the stress-energy tensor 
[17]. In any case, we hope to discuss the situation with matter terms elsewhere. 

\medskip

  As stated, these results do not hold for spacetimes containing black holes, 
or for any spacetimes containing singularities which propagate to ${\mathcal I}$. 
As a specific example, it is well-known that the AdS-Kerr spacetime has conformal 
infinity ${\mathcal I}$ given by a finite-time region $I\times S^{n-1}$ of the 
Einstein cylinder ${\mathbb R}\times S^{n-1}$. Not all symmetries of $S^{n-1}$ 
extend to symmetries of the AdS-Kerr solution, so there are timelike Killing fields 
on ${\mathcal I}$ which do not extend to Killing fields on the Kerr-AdS solution. 
More generally, smooth global initial data on a space-like slice $\Sigma$ may evolve 
to the future (or past) and eventually form a black hole, for example the AdS Kerr 
metric. Such solutions will be asymptotically stationary to the future (or past) 
near ${\mathcal I}$ within their domain of outer communication, but are not 
themselves globally stationary; see also Remark 3.1. 

\medskip

  The contents of the paper are briefly as follows. In \S 2, we present some basic 
background material and elementary results needed to establish the main results. 
These results are then proved in \S 3. In \S 4, we conclude with further discussion 
and interpretation of the results. 

\bigskip

   I would like to thank Piotr Chru\'sciel, Gary Gibbons, Gregory Jones for discussions 
and correspondence on issues related to this work. Particular thanks to Kostas 
Skenderis and Bob Wald for numerous comments and criticisms on an earlier version 
of this work. This project began during a stay at the Newton Institute, Cambridge 
in Fall 05, and I am grateful to the Institute for its hospitality.

\section{ Background Material.}
\setcounter{equation}{0}

 Let $\Sigma$ be a compact $n$-manifold with boundary $\partial\Sigma$ 
and $M = {\mathbb R}\times \Sigma$, where ${\mathbb R}$ represents the time 
direction. We consider geodesically complete solutions $(M, g)$ of the 
vacuum Einstein equations (1.1) which are asymptotically simple. 
Asymptotic simplicity is equivalent to the existence of a (reasonably 
smooth) conformal completion. Thus, let $\widetilde \rho$ be a defining 
function for the boundary $\partial M = {\mathcal I}  = {\mathbb R} \times 
\partial\Sigma$, i.e.~$\widetilde \rho$ is a coordinate function for 
$\partial M$ in $M$, vanishing on $\partial M$. One then requires that 
the conformally equivalent (unphysical) metric 
\begin{equation} \label{e2.1}
\widetilde g = \widetilde \rho^{2}g 
\end{equation}
extends at least $C^{2}$ to ${\mathcal I}$; (a stronger smoothness condition 
will be required below). It is easy to see that this implies that 
the sectional curvatures $K$ of $g$ satisfy $|K+1| = 
O(\widetilde \rho^{2})$, so that the metric $g$ locally approaches the AdS 
metric to order $O(\widetilde \rho^{2})$ near ${\mathcal I}$; for this 
reason, asymptotically simple spacetimes are also often called 
asymptotically locally AdS. The space ${\mathcal I}$ is called conformal 
infinity. 

 Let 
\begin{equation} \label{e2.2}
\gamma  = \widetilde g|_{{\mathcal I}} 
\end{equation}
be the metric induced on ${\mathcal I}$. It is well-known that causality 
arguments imply that $\gamma$ is a Lorentz metric on ${\mathcal I}$. 
Different choices of defining function $\widetilde \rho$ lead to 
conformally equivalent metrics $\gamma $ on ${\mathcal I}$, so that only 
the conformal structure $[\gamma]$ on ${\mathcal I}$ is uniquely 
determined by $(M, g)$. 

 If $(M, g)$ is asymptotically simple, then it is standard, (and easily seen), 
that each choice of boundary metric $\gamma \in [\gamma]$ determines a unique 
geodesic defining function $\rho$, for which the integral curves of 
$\bar \nabla \rho$ in the compactified metric 
\begin{equation} \label{e2.3}
\bar g = \rho^{2}g 
\end{equation}
are (spacelike) geodesics orthogonal to ${\mathcal I}$. In the following, 
we work only with such geodesic compactifications. (The integral curves 
of $\nabla \log \rho$ are also geodesics with respect to the metric $g$). 
The metric $g$ splits in the $\rho$-direction, so that 
\begin{equation} \label{e2.4}
g = \rho^{-2}(d\rho^{2} + g_{\rho}), 
\end{equation}
where $g_{\rho}$ is a curve of Lorentz metrics on the level sets 
$S(\rho)$ of $\rho$. As $\rho \rightarrow 0$, $g_{\rho} \rightarrow 
\gamma \equiv g_{(0)}$. The Fefferman-Graham expansion [20] is the 
expansion of the curve $g_{\rho}$ in a Taylor-type series in $\rho$. 
The exact form of the expansion depends on whether $n$ is even or odd. 
If $n$ is odd, then one has
\begin{equation} \label{e2.5}
g_{\rho} \sim  g_{(0)} + \rho^{2}g_{(2)} + \cdots + \rho^{n-1}g_{(n-1)} + 
\rho^{n}g_{(n)} + \cdots
\end{equation}
while if $n$ is even, 
\begin{equation} \label{e2.6}
g_{\rho} \sim  g_{(0)} + \rho^{2}g_{(2)} + \cdots + \rho^{n-1}g_{(n-1)} + 
\rho^{n}\log \rho \, {\mathcal H}  + \rho^{n}g_{(n)}+ \cdots
\end{equation}
Below order $n$, the expansions are even in powers of $\rho$, and all 
coefficients $g_{(2k)}$, $2k < n$, as well as the coefficient ${\mathcal H}$ 
in (2.6) are determined by the boundary metric $\gamma  = g_{(0)}$ and 
its derivatives up to order $2k$, (respectively $n$); thus they do not 
depend on the particular bulk Einstein metric $g$. The series (2.5) is a 
formal power series, in powers of $\rho$, while the series (2.6) is a 
series in powers of $\rho$ and log $\rho$, (i.e.~a polyhomogeneous series). 

 The coefficient $g_{(n)}$ is formally undetermined; thus it is not 
determined by $\gamma$, and depends on the bulk metric $g$. All the 
remaining (higher order) coefficients in the series (2.5)-(2.6) are 
completely determined by the data
\begin{equation} \label{e2.7}
(g_{(0)}, g_{(n)}), 
\end{equation}
so that these terms determine the formal expansion of the metric 
$\bar g$, and hence $g$, near ${\mathcal I}$. Note that the terms in the 
expansion (2.5)-(2.6) depend on the choice of boundary metric 
$\gamma\in [\gamma]$. A conformal change of $\gamma$ will cause a 
change in the geodesic defining function $\rho$, and hence a change in 
the coefficients. Transformation formulae for these coefficients are 
given in [17], [21] for instance. 

 The boundary metric $g_{(0)}$ is (formally) free, in the sense that 
the Einstein equations in a neighborhood of ${\mathcal I}$ impose no 
conditions or constraints on $g_{(0)}$. Similarly, the 
transverse-traceless part of $g_{(n)}$ is formally free; however, 
the Einstein equations do impose constraints on the divergence and 
trace of $g_{(n)}$. Thus
\begin{equation} \label{e2.8}
\delta g_{(n)} = r_{(n)}, \ \ {\rm and} \ \  tr g_{(n)} = s_{(n)}, 
\end{equation}
where the divergence $\delta$ and trace are taken with respect to 
$g_{(0)}$. The terms $r_{(n)}$ and $s_{(n)}$ may be explicitly computed 
from the boundary metric $g_{(0)}$ and its derivatives up to order $n$, 
cf. [17]. When $n$ is odd, one has $r_{(n)} = s_{(n)} = 0$. We 
will call the equations (2.8) the constraint equations on ${\mathcal I}$. 
They arise from the Gauss and Gauss-Codazzi equations on ${\mathcal I}$ in 
the geodesic gauge (2.4). 

\medskip

 Throughout the following, we assume that the metric $g$ is 
asymptotically simple in the sense that the expansion (2.5) or (2.6) 
exists to order $n$, so that 
\begin{equation} \label{e2.9}
g_{\rho} = g_{(0)} + \rho^{2}g_{(2)} + \cdots + \rho^{n-1}g_{(n-1)} + 
\rho^{n}g_{(n)} + o(\rho^{n}), 
\end{equation}
or, 
\begin{equation} \label{e2.10}
g_{\rho} = g_{(0)} + \rho^{2}g_{(2)} + \cdots + \rho^{n-1}g_{(n-1)}+ 
\rho^{n}\log \rho \, {\mathcal H}  + \rho^{n}g_{(n)} + o(\rho^{n}), 
\end{equation}
where $o(\rho^{n})/\rho^{n} \rightarrow 0$ as $\rho \rightarrow 0$. 

 We point out that if the free data $(g_{(0)}, g_{(n)})$ are 
real-analytic and satisfy the constraints (2.8), then a result of 
Kichenassamy [22] shows that the (formal) series expansion (2.5) or 
(2.6) converges to an actual solution $g$ of the Einstein equations 
near ${\mathcal I}$. Of course, since the series is uniquely determined by 
the data (2.7), this solution is unique. (It is mathematically an open 
question whether there could exist other solutions with the same data 
(2.7) which don't have convergent expansions; this seems very unlikely 
however). 

 Throughout most of the paper, we only consider boundary metrics $\gamma  = 
g_{(0)}$ which are globally stationary on ${\mathcal I}$. Thus, there is a complete, 
timelike Killing field $Z$, generating a free isometric ${\mathbb R}$-action 
on $({\mathcal I}, \gamma)$, so that 
\begin{equation} \label{e2.11}
{\mathcal L}_{Z}\gamma  = 0. 
\end{equation}
Let $\pi: {\mathcal I} \rightarrow S$ be the projection onto the orbit 
space of the ${\mathbb R}$-action. The ${\mathbb R}$-bundle $\pi$ is 
trivial, $M \simeq {\mathbb R}\times S$, and with respect to a fixed 
trivialization determined by a global time function $t$ on ${\mathcal I}$, 
the metric $\gamma$ may be written globally in the form
\begin{equation} \label{e2.12}
\gamma  = -N^{2}(dt+\theta )^{2} + \pi^{*}g_{S}, 
\end{equation}
where $\theta$ is a connection 1-form on the bundle $\pi$, $N^{2} = 
-\gamma (Z,Z) > 0$ and $g_{S}$ in the Riemannian metric induced on the 
orbit space $S$ by $\gamma$. 

 We require that $\gamma$ is $C^{n+1,\alpha}$ smooth, but otherwise 
impose no other conditions on $\gamma$; it need not satisfy any 
equations or have any other symmetries. If $\gamma$ does have other 
symmetries, i.e.~the metric $g_{S}$ also admits Killing fields, the 
timelike Killing field $Z$ will not be unique. The results to follow hold 
for any fixed choice of $Z$. 

 It follows that if $g$ is an asymptotically simple solution of the 
Einstein equations with boundary metric $\gamma$, then the determined 
coefficients in (2.5)-(2.6), i.e.~$g_{(k)}$ with $k \leq n-1$, and the 
logarithmic coefficient ${\mathcal H}$, are also invariant under the isometric 
action generated by $Z$, 
\begin{equation} \label{e2.13}
{\mathcal L}_{Z}g_{(k)} = 0, \ \ {\mathcal L}_{Z}{\mathcal H}  = 0. 
\end{equation}

\begin{definition} \label{d2.1}
{\rm The space ${\mathcal E} = {\mathcal E}_{\gamma}$ is the space of all geodesically 
complete, asymptotically simple solutions of the vacuum Einstein equations which 
have the fixed stationary metric $\gamma$ in (2.12) as boundary metric and for which 
the expansions (2.9)-(2.10) hold to order $n+1+\alpha$, $\alpha \in (0,1)$. One may 
define a natural ${\mathcal C}^{n+1,\alpha}$ polyhomogeneous topology on 
${\mathcal E}$. }
\end{definition}

  Let $t$ be a fixed smooth global time function on $M$, which restricts to the 
time function $t$ above on ${\mathcal I}$. This gives a trivialization $M = 
{\mathbb R}\times \Sigma$, with fibers $\Sigma_{t}$ given by the level sets 
of $t$. With respect to this foliation $\Sigma_{t}$, the metric $g$ may be 
written in local coordinates $(t, x^{i})$ as
\begin{equation} \label{e2.14}
g = -u^{2}dt^{2} + g_{ij}(dx^{i} + \xi^{i}dt)(dx^{j} + \xi^{j}dt), 
\end{equation}
where $(N, \xi)$ is the lapse-shift vector with respect to $\partial_{t}$. 
Since the boundary metric of $g$ with respect to $\rho$ is $\gamma$, one has 
$u\rho  \rightarrow N$, for $N$ as in (2.12), as $\rho \rightarrow 0$. 

\medskip

 Let ${\mathcal C}$ be the space of solutions of the constraint equations 
on a given $\Sigma$, say $\Sigma_{0}$ as above, induced by the global 
metrics $g\in{\mathcal E}$. Thus, an element in ${\mathcal C}$ is a triple 
$(\Sigma, g_{0}, K)$, where $g_{0}$ is a complete, conformally compact 
metric on $\Sigma$, and $K$ is a symmetric bilinear form, (the 2nd 
fundamental form), with $K = g_{0} + O(\rho^{2})$. The data $(g_{0}, K)$ 
satisfy the vacuum constraint equations
\begin{equation} \label{e2.15}
R - |K|^{2} + \kappa^{2} = n(n-1), 
\end{equation}
$$\delta K + d\kappa = 0, $$
where $\kappa = tr K$ is the mean curvature. 

 Note that ${\mathcal C}$ does not consist of all solutions of the 
constraint equations (2.15); many solutions of (2.15) will only give 
rise to local-in-time solutions of the vacuum equations (1.1). The 
space ${\mathcal C}$ may be given a ${\mathcal C}^{n+1,\alpha}\times 
{\mathcal C}^{n,\alpha}$ polyhomogeneous topology induced from the 
topology of ${\mathcal E}$. 

 Given the spacelike foliation $\Sigma_{t}$ above, let $g_{t}$ be the 
metric induced by $g\in{\mathcal E}$ on $\Sigma_{t}$. Choosing the zero-shift 
gauge gives a diffeomorphism $\phi_{t}: \Sigma = \Sigma_{0} \rightarrow  
\Sigma_{t}$, and we will let $g_{t}$ also denote the induced metric 
$\phi_{t}^{*}(g_{t})$ on $\Sigma$. Thus, the Einstein flow from 
$\Sigma$ to any $\Sigma_{t}$ defines a flow, denoted 
\begin{equation} \label{e2.16}
g_{0} \rightarrow  g_{t}, \ K_{0} \rightarrow K_{t} 
\end{equation}
on the constraint space ${\mathcal C}$. 

\medskip

 Next we discuss briefly the linearized Einstein equations. The 
linearization of the vacuum equations (1.1) at $g \in {\mathcal E}$ is 
given by 
\begin{equation} \label{e2.17}
D^{*}Dh - 2Rh + \delta^{*}\delta\bar h = 0. 
\end{equation}
Here $h$ is a symmetric bilinear form, $\bar h = h - 
\frac{1}{2}trh\, g$, $D^{*}D = tr D^{2}$ is the wave operator and $R$ 
is the curvature tensor acting on symmetric bilinear forms; all metric 
quantities in (2.17) are with respect to the background metric $g$. 
Altering $h$ by infinitesimal diffeomorphisms $h \rightarrow  h + 
\delta^{*}X$, it is well-known that one can solve (2.17) in the 
transverse gauge $\delta \bar h = 0$, by solving the coupled system
\begin{equation} \label{e2.18}
D^{*}D(h+\delta^{*}X) - 2R(h+\delta^{*}X) = 0, 
\end{equation}
\begin{equation} \label{e2.19}
\delta (\bar h +\bar{\delta}^{*}X) = 0, 
\end{equation}
in the variables $(h, X)$, cf.~[23]. 

 Consider the initial boundary value problem for (2.18):
\begin{equation} \label{e2.20}
D^{*}Dh - 2Rh = 0, \ h|_{\Sigma} = h_{0}, \ \nabla_{\partial_{t}}h = h_{1}, 
\ {\rm and} \  h|_{{\mathcal I}} = h_{(0)}, 
\end{equation}
where the boundary data $h_{(0)} \in  C^{\infty}({\mathcal I})$ and the 
initial data $h_{0}, h_{1}$ are $C^{\infty}$ polyhomogeneous on $\Sigma 
 = \Sigma_{0}$ up to the boundary. We assume $h_{0}$, $h_{1}$ and 
$h_{(0)}$ match in a smooth polyhomogeneous sense at the corner 
$\partial\Sigma$. In this generality, it is not known if there is a 
global $C^{\infty}$ polyhomogeneous solution $h$ of (2.20) defined on 
all of $M$.\footnote{The logarithmic terms in the expansion of $h$ at 
${\mathcal I}$ may propagate into the bulk of $M$ and could, apriori, 
lead to singularities of the solution in the bulk. I am grateful to 
P. Chru\'sciel for discussions on this point.} However, it suffices 
for our purposes, (namely regarding the unique continuation property), 
that one has a smooth solution in the interior of the domain of dependence 
$D(\Sigma) \subset M$, which is $C^{n+1,\alpha}$ polyhomogeneous smooth 
at the boundary $\partial \Sigma \subset {\mathcal I}$. The existence of 
such solutions follows from standard energy estimates, cf.~[24] for 
instance. Such energy estimates are carried out in the Sobolev spaces 
$H^{s}$; note that Sobolev embedding gives $C^{n+1,\alpha} \subset H^{s}$, 
for $s > \frac{3}{2}n + 2$. In fact, it suffices for our purposes to know 
that there exist solutions to the linearization of the constraint equations 
(2.15) which are $C^{n+1,\alpha}$ polyhomogeneous at $\partial \Sigma$; this 
follows from the work of [25].  

 A standard identity (Weitzenbock formula) gives 
$\delta\bar \delta^{*}X = \frac{1}{2}D^{*}DX - Ric X$, so that (2.19) 
is equivalent to the vector system 
\begin{equation} \label{e2.21}
{\tfrac{1}{2}}D^{*}DX - Ric X = -\delta\bar h, 
\end{equation}
which is of the same form as (2.20). Thus, as above, one may solve the 
initial boundary value problem for (2.21) within $D(\Sigma)$. 

 The symmetric bilinear forms $h$ satisfying
\begin{equation} \label{e2.22}
D^{*}Dh - 2Rh = 0, 
\end{equation}
$$\delta\bar h = 0, $$
which satisfy $h_{(0)} = 0$ on ${\mathcal I}$ may be viewed as defining 
the formal tangent space $T_{g}{\mathcal E}$ to ${\mathcal E}$, (modulo 
diffeomorphisms). However, this is formal; no claim is made that 
${\mathcal E}$ is a manifold, corresponding to the linearization stability 
of the Einstein equations within the space ${\mathcal E}$. Moreover, 
solutions of (2.22) with 
$$h_{(0)} \neq 0 \ {\rm on} \  {\mathcal I} , $$
are certainly not in the formal tangent space $T{\mathcal E}$, since these 
deformations don't preserve the boundary metric. 

 We now define precisely the unique continuation property (iii) of 
Theorem 1.1. This is the statement that, for $g \in {\mathcal E}$, any 
solution $h$ of the linearized Einstein equations (2.22) which vanishes 
to infinite order at ${\mathcal I}$ is necessarily zero. For solutions $h$ 
of (2.22) which are $C^{\infty}$ polyhomogeneous up to ${\mathcal I}$, this 
is clearly equivalent to the statement that if $h$ has zero Cauchy data 
on ${\mathcal I}$, i.e.~if 
\begin{equation} \label{e2.23}
h_{(0)} = 0 \ {\rm and} \  h_{(n)} = 0, 
\end{equation}
then 
\begin{equation} \label{e2.24}
h \equiv  0. 
\end{equation}
This follows from the properties of expansions the (2.5)-(2.6); the 
conditions (2.23) imply that the formal series solution of (2.22) 
vanishes. Note that it suffices to have the unique continuation 
property at a cut $\partial\Sigma$ of ${\mathcal I}$, i.e. within the 
domain of dependence $D(\Sigma)$. For if $h$ vanishes to 1st order on 
$\Sigma$ and $h_{(0)} = 0$, then $h \equiv 0$, by uniqueness of 
solutions to the initial boundary value problem (2.22). 

\medskip

 Next we return to the expansions (2.5)-(2.6). The undetermined term 
$g_{(n)}$ is closely related to the stress-energy tensor on ${\mathcal I}$, 
and is an important feature of the AdS/CFT correspondence. Thus, 
as shown in [16], [18], there is a symmetric bilinear form $r_{(n)}$, 
determined by the boundary metric $\gamma$ and its derivatives up to 
order $n$, such that the form 
\begin{equation} \label{e2.25}
\tau_{(n)} = g_{(n)} - r_{(n)}, 
\end{equation}
is divergence-free with determined trace, i.e.
\begin{equation} \label{e2.26}
\delta_{\gamma}\tau_{(n)} = 0, \ \ tr_{\gamma}\tau_{(n)} = a_{(n)}. 
\end{equation}
The term $\tau_{(n)}$ is obtained by a covariant (intrinsic) renormalization 
of the Brown-York quasi-local stress-energy tensor. Via the AdS/CFT 
correspondence, $\tau_{(n)}$ corresponds to the expectation value of the 
stress-energy tensor of the QFT dual to $(M, g)$. The term $a_{(n)}$ is 
proportional to the conformal anomaly [26], and is determined by the boundary 
metric $\gamma$. When $n$ is odd, $r_{(n)} = 0$, so $\tau_{(n)} = g_{(n)}$ and 
$a_{(n)} = 0$. 

 It is important to note that the construction of $\tau_{(n)}$ is 
background-independent; there is no normalization with respect to a 
background ``standard'' solution. Indeed, there are no such standard 
background solutions to which $(M, g)$ can be compared in the generality 
of the current discussion. For instance, apriori, there may not be any 
stationary metric on $M$ with conformal infinity $({\mathcal I}, \gamma)$. 
A more recent and efficient construction of the stress-energy tensor 
$\tau_{(n)}$ is given in [18], [19], cf.~also [27], [28]. 

 Given $g \in {\mathcal E}$, the (holographic) mass, cf. [16], [17], of the cut 
$\partial\Sigma_{t} \subset  {\mathcal I}$ is defined by 
\begin{equation} \label{e2.27}
m_{\partial\Sigma_{t}} = \int_{\partial\Sigma_{t}}\tau_{(n)}(Z, \nu )dV_{\gamma}, 
\end{equation}
where $\nu$ is the future unit normal and $Z$ is the (future-oriented) Killing 
field on ${\mathcal I}$. By its definition, this mass is independent of spacelike 
hypersurfaces $\Sigma_{t} \subset (M, g)$ giving the same cut $\partial\Sigma_{t}$ 
at ${\mathcal I}$. Both the holographic mass $m$ and the stress-energy tensor 
$\tau_{(n)}$ depend on the choice of boundary metric, (or equivalently 
on the choice of defining function $\widetilde \rho$). However, as noted above, 
the boundary metric $\gamma$ is chosen to be the fixed stationary metric (2.12). 

 Since $\gamma$ is stationary, a standard application of (2.26) and the 
divergence theorem implies that 
\begin{equation} \label{e2.28}
m = m_{\partial\Sigma_{t}}, 
\end{equation}
is independent of the cut $\partial\Sigma_{t}$, (and hence $t$), so that 
the mass depends only on the solution $(M, g) \in {\mathcal E}$, (given 
the fixed choice of boundary metric), and the choice of timelike Killing 
field $Z$ on $({\mathcal I}, \gamma)$. In other words, the mass is conserved. 
If $\gamma$ has a larger space of timelike Killing fields, one may consider 
the mass (2.27) defined with respect to each choice, or take the ``canonical'' 
Killing field with zero angular velocity, cf. [19]. For the rest of the paper, 
$Z$ denotes any fixed timelike Killing field. 

 There have been numerous definitions of mass and other conserved quantities 
for asymptotically AdS and asymptotically locally AdS spacetimes, cf.~[29], 
[30], [31], for example; cf.~[32] for an overview. In the generality considered 
here, the holographic mass in the only suitable definition, again since it is 
background-independent and covariant or intrinsic to the metric $(M, g)$, given 
a fixed choice of boundary metric. Many of the various definitions of mass have 
recently been shown to be equivalent to the holographic mass, cf.~[33] and in 
particular [19] for a very clear analysis. 

\medskip

  Next, consider variations of the mass with respect to variations of 
the metric $g \in {\mathcal E}$. Suppose then that $h$ is an infinitesimal 
Einstein deformation, with induced boundary variation $h_{(0)}$. It is 
convenient to put the variation $h$ in the geodesic gauge for a fixed 
geodesic defining function $\rho$, so that $h$ respects the splitting 
(2.4), i.e.~$h_{0\alpha} = 0$. This can always be accomplished by a suitable 
gauge transformation, (i.e.~infinitesimal diffeomorphism). Since the curve 
$g_{u} = g + uh$ is Einstein to 1st order in $u$, the constraint equations 
(2.26) give
\begin{equation} \label{e2.29}
\delta' (\tau_{(n)}) + \delta\sigma_{(n)} = 0, 
\end{equation}
where $\delta'$ is the variation of the divergence $\delta$ in the 
direction of the variation of the boundary metric $\gamma$ and 
$\sigma_{(n)} = \tau_{(n)}'  = h_{(n)} - r_{(n)}'$. Suppose the 
boundary metric $\gamma$ is kept fixed, i.e. $h_{(0)} = 0$, so that, 
formally, $h\in T_{g}{\mathcal E}$. Then
\begin{equation} \label{e2.30}
\delta\sigma_{(n)} = 0. 
\end{equation}
By definition, one has
\begin{equation} \label{e2.31}
dm_{\partial\Sigma_{t}}(h) = \frac{dm_{u,t}}{du}|_{u=0} = 
\int_{\partial\Sigma_{t}}\frac{d}{du}(\tau_{(n)})_{g+uh}(Z, \nu )dV = 
\int_{\partial\Sigma_{t}}\sigma_{(n)}(Z, \nu )dV, 
\end{equation}
again since the boundary metric is fixed. Thus, via (2.30) and the 
divergence theorem, $dm_{\partial\Sigma}$ is independent of the cut 
$\partial\Sigma$, i.e.~is conserved, for any fixed $h$ vanishing on 
${\mathcal I}$. So for such variations, both the mass as well as its 
variation, are conserved. 

\medskip

 The following discussion will be important in the proof of Theorem 1.1 
in \S 3. For an arbitrary complete Lorenztian boundary metric $\gamma $ 
on ${\mathcal I}  = {\mathbb R} \times \partial\Sigma$, let ${\mathcal T}$ be the 
space of (smooth) symmetric bilinear forms on ${\mathcal I}$ which satisfy the 
constraint equations (2.26) with respect to $\gamma$; when $n$ is odd, these 
are the transverse-traceless (TT) forms. Thus ${\mathcal T}$ is naturally an 
affine bundle  
$$\pi : {\mathcal T}  \rightarrow  Met({\mathcal I} ) $$
over the space of Lorentz metrics on ${\mathcal I}$. An element $(\gamma, \tau)$, 
$\tau \in {\mathcal T}_{\gamma}$, in ${\mathcal T}$ then defines a unique 
formal series solution of the Einstein equations (1.1), defined near 
${\mathcal I}$. This follows immediately from the discussion following 
(2.5)-(2.6). If the pair $(\gamma, \tau)$ are real-analytic, the formal 
series converges to an actual solution $g$ of the Einstein equations (1.1), 
again defined in a neighborhood of ${\mathcal I}$. 

\begin{proposition} \label{p2.2}
At any $(\gamma, \tau) \in {\mathcal T}$, the map $\pi$ is a submersion, 
i.e.~its derivative is surjective, and so $\pi$ is locally surjective. 
\end{proposition}

{\bf Proof:} Given $(\gamma, \tau) \in {\mathcal T}$, one needs to show that 
for any variation $\gamma' = h_{(0)}$ of $\gamma$, there exist 
solutions $\tau'$ of the linearized constraint equations (2.26), i.e. 
\begin{equation} \label{e2.32}
\delta'\tau  + \delta\tau'  = 0, 
\end{equation}
$$tr'\, \tau  + tr\, \tau'  = a_{(n)}'. $$
Let $\sigma  = \tau'$, so that it suffices to solve
\begin{equation} \label{e2.33}
\delta\sigma  = \phi_{1}, \ \ tr \sigma  = \phi_{2}, 
\end{equation}
for arbitary $\phi_{1}, \phi_{2}$. Consider for instance $\sigma$ of 
the form $\sigma  = \bar \delta^{*}V + f\gamma$, where $V$ is a 
vector field. Then (2.33) becomes the system 
$\delta\bar \delta^{*}V - df = \phi_{1}$, 
$tr(\bar \delta^{*}V) + nf = \phi_{2}$, so that
$$\delta\bar \delta^{*}V + {\tfrac{1}{n}}dtr \bar \delta^{*}V  = 
\phi_{1} - {\tfrac{1}{n}}d\phi_{2}. $$
One has $\delta\bar \delta^{*}V = \frac{1}{2}D^{*}DV - Ric V$, while 
tr $\bar \delta^{*}V = (\frac{n}{2}-1)\delta V$. Hence, the 
equation above is equivalent to
\begin{equation} \label{e2.34}
{\tfrac{1}{2}}D^{*}DV + ({\tfrac{1}{2}-\tfrac{1}{n}})d\delta V - RicV = 
\phi_{1} - {\tfrac{1}{n}}d\phi_{2}. 
\end{equation}
This is a linear hyperbolic system for $V$ on ${\mathcal I}$ with 
$\partial\Sigma$ a compact Cauchy surface. It is standard that the Cauchy 
problem for (2.34) has global solutions $V$ on $({\mathcal I}, \gamma )$ 
with arbitrary initial data. Given $V$, one may then solve the trace equation 
above to obtain $f$ and the resulting pair $\sigma$ satisfying (2.32). 

{\endproof}

 Just as before with pairs $(\gamma, \tau)$ satisfying the constraint 
equations (2.26), the space of pairs $(h_{(0)}, \sigma)$, for $\sigma$ 
satisfying (2.32), corresponds exactly to the space of formal series 
solutions of the linearized Einstein equations defined near ${\mathcal I}$, 
(in the geodesic gauge). Again if $(h_{(0)}, \sigma)$ are real-analytic, 
the formal series converges to an actual solution of the linearized Einstein 
equations defined in a neighborhood of ${\mathcal I}$. 

 For a given $h_{(0)}$, the space of solutions $\sigma$ of (2.32) is an 
affine space ${\mathcal F}_{h_{(0)}}$. Observe that the space of global 
solutions, (or solutions defined in $D(\Sigma)$), of the Einstein equations 
linearized at $g$ with induced boundary variation $h_{(0)}$ is also an 
affine space ${\mathcal G}_{h_{(0)}}$, parameterized by the Cauchy data 
$h_{0}, h_{1}$ on a spacelike hypersurface $\Sigma \subset (M, g)$. Clearly,
$${\mathcal G}_{h_{(0)}} \subset  {\mathcal F}_{h_{(0)}}. $$
We do not know if the two spaces ${\mathcal G}_{h_{(0)}}$, ${\mathcal F}_{h_{(0)}}$ 
actually coincide, although there seems to be no compelling reason for this 
to be the case. 

\begin{lemma} 
Suppose $\gamma$ is globally stationary. Then for any smooth variation $h_{(0)}$ of 
compact support on ${\mathcal I}$, there exist solutions $\sigma$ to (2.32) on 
${\mathcal I}$ which are uniformly bounded, in that 
\begin{equation} \label{e2.35}
|\sigma|_{C^{1}} \leq  K, 
\end{equation}
where $K$ depends only on $(\gamma , h_{(0)})$ and is independent of $t$. 
\end{lemma}
{\bf Proof:} We will constuct a specific bounded solution, although it is 
clear that there will be many other possibilities. Let $e_{\alpha}$, $0 \leq 
\alpha \leq n$, be a local orthonormal framing of $({\mathcal I}, \gamma)$, 
with $e_{0} = T = Z/|Z|$, so that $\{e_{i}\}$ are tangent to 
the orbit space $S$ of ${\mathcal I}$. At a given point $p \in S$, assume that 
$\nabla_{e_{i}}e_{j}(p) = 0$, where here, and only here, the covariant derivative 
is on $(S, \gamma_{S})$. Then, at $p$, $-\delta\sigma  = (\nabla_{T}\sigma )(T) + 
(\nabla_{e_{i}}\sigma )(e_{i}) = \nabla_{T}\sigma (T) - \sigma (\nabla \log N) + 
\nabla_{e_{i}}\sigma (e_{i})$, since $\langle \nabla_{e_{i}}e_{i}, T\rangle = 0$. 
(The last equation follows since $T$ is colinear to a Killing field orthogonal 
to $e_{i}$). Let $\sigma^{H}$ be the horizontal part of $\sigma$, $\sigma^{H} = 
\sigma - \sigma (T)\cdot T$.  

 For simplicity, suppose $\sigma (T, e_{i}) = 0$, for all $i$. Then a 
simple calculation shows that the first equation in (2.33) becomes
\begin{equation}\label{e2.36}
-(\nabla_{T}\sigma (T))(T) = \phi (T),
\end{equation}
\begin{equation}\label{e2.37}
\delta^{H}(N\sigma^{H}) = N\phi^{H},
\end{equation}
where $\phi^{H} = \phi - \phi (T)T$ is the horizontal projection of 
$\phi$ and $\delta^{H}$ is the divergence operator on the orbit space 
$(S, g_{S})$. Since $h_{(0)}$ has compact support, so does $\phi$. 

   The equations (2.36)-(2.37) are uncoupled. The first equation 
may be solved directly by integration along the integral curves of $T$. 
Since $\phi (T)$ has compact support, $\sigma (T,T)$ remains uniformly 
bounded in the sense of (2.35). 

 For the second equation, let ${\mathbb S}^{2}(S)$ be the space of symmetric 
bilinear forms on $S$. Then ${\mathbb S}^{2}(S) = Im (\delta^{H})\oplus Ker 
((\delta^{H})^{*})$, and the second summand corresponds to the space of 
Killing fields on $(S, \gamma_{S})$. Suppose first the orbit space 
$(S, g_{S})$ has no Killing fields, so that the operator $\delta^{T}: 
{\mathbb S}^{2}(S) \rightarrow  \Omega^{1}(S)$ is surjective. Then (2.37) 
admits (many) solutions, with $\sigma^{H}$ satisfying (2.35). 

  If $(S, \gamma_{S})$ admits Killing fields, then by linearity it suffices 
to solve
\begin{equation} \label{e2.38}
\delta \sigma = X_{t},
\end{equation}
where, for each fixed $t$, $X_{t}$ is a Killing field on $S$ and $X_{t}$ has 
compact support on ${\mathcal I}$. Consider first the equation
$$\delta \sigma = f(t)X,$$
where $X$ is a fixed Killing field on $S$. Setting $\sigma = v(t)T\cdot X$, 
a simple computation gives 
$$-\delta(v(t)T\cdot X) = v'(t)X + \nabla_{X}T + \nabla_{T}X.$$
Straightforward computation using the Killing properties shows that 
$\nabla_{X}T + \nabla_{T}X$ is orthogonal to $T$ and the space $\chi(S)$ 
of Killing fields on $S$, and hence $\nabla_{X}T + \nabla_{T}X 
\in Im(\delta^{H})$. Setting $v' = -f$ and using linearity then shows 
that the first equation in (2.33) is solvable with bounded $\sigma$. 
A general $X_{t}$ as in (2.38) has the form $X_{t} = \sum f_{i}(t)X_{i}$, 
where $X_{i}$ is a basis of $\chi(S)$ and so the result for the first 
equation in (2.33) again follows by linearity. 

  To solve the second equation in (2.33), by linearity it suffices to 
solve $\delta \sigma = 0$ and $tr \sigma = \phi_{2}$. It is clear that 
this system has many $C^{1}$ bounded solutions. 

{\endproof}

 It is not clear in this generality that there exists a solution $h$ of 
the linearized Einstein equations, with $h|_{{\mathcal I}} = h_{(0)}$ of 
compact support for which $\sigma_{(n)}$ satisfies (2.35). 

\section{Proofs of the Results.}
\setcounter{equation}{0} 

 The main tool used in the proof of Theorems 1.1 and 1.2 is the 
following identity, proved in [10]; for completeness, the proof is also 
given in the Appendix. Let $X$ be any Killing field on $({\mathcal I}, \gamma)$ 
and let $\tau$ be any smooth symmetric bilinear form on ${\mathcal I}$, 
satisfying the constraint equations (2.26). (More precisely, only the 
divergence-free condition in (2.26) is needed). If $h_{(0)}$ is any 
variation of $\gamma$ of compact support on ${\mathcal I}$, then
\begin{equation} \label{e3.1}
\int_{{\mathcal I}}\langle {\mathcal L}_{X}\tau, h_{(0)}\rangle dV = 
-2\int_{{\mathcal I}}\langle \delta'\tau, X\rangle dV, 
\end{equation}
where as in (2.29), $\delta'$ is the variation of the divergence $\delta = 
\delta_{\gamma}$ in the direction $h_{(0)}$. (A simple modification 
of (3.1) also holds for conformal Killing fields, cf.~(A.6)). 

 The relation (3.1) holds in particular for $\tau  = \tau_{(n)}$, 
where $\tau_{(n)}$ is the stress-energy tensor associated to a 
solution $g \in {\mathcal E}$. Further, by Proposition 2.2 or Lemma 2.3, 
the equation 
\begin{equation} \label{e3.2}
\delta'\tau_{(n)} = -\delta\sigma , 
\end{equation}
is always solvable, for some symmetric bilinear form $\sigma$ on 
${\mathcal I}$. Clearly $\sigma$ is determined only up to a divergence-free 
symmetric bilinear form. 

 Choose $X = Z$ a timelike Killing field on $({\mathcal I}, \gamma)$ and 
suppose the variation $h_{(0)}$ of the boundary metric $\gamma$ has 
compact support; let $\partial\Sigma^{\pm}$ be any two cuts of 
${\mathcal I}$ which enclose supp $h_{(0)}$. Suppose also that $\sigma  = 
\sigma_{(n)}$ arises from a (global) solution $h$ to the linearized 
Einstein equations, i.e.~$\sigma_{(n)} = \tau_{(n)}' = h_{(n)} - r_{(n)}'$, 
(cf. (2.29)). Then (3.1) and the divergence theorem applied to (3.2) give
\begin{equation} \label{e3.3}
\int_{{\mathcal I}}\langle {\mathcal L}_{Z}\tau_{(n)}, h_{(0)}\rangle dV = 
-2\int_{\partial\Sigma^{\pm}}\sigma_{(n)}(Z, \nu )dV = -2[dm^{+}(h) 
- dm^{-}(h)], 
\end{equation}
where $dm^{\pm}$ is the variation of the holographic mass (2.27) at 
$\partial\Sigma^{\pm}$, cf. also (2.31). It is clear that this formula 
holds for any formal solution $h_{f} = (h_{(0)}, \sigma)$ of the linearized 
constraint equations (2.32) or (3.2): 
\begin{equation} \label{e3.4}
\int_{{\mathcal I}}\langle {\mathcal L}_{Z}\tau_{(n)}, h_{(0)} \rangle dV = 
-2\int_{\partial\Sigma^{\pm}}\sigma (Z, \nu )dV = -2[dm^{+}(h_{f}) - 
dm^{-}(h_{f})]. 
\end{equation}

\medskip

 {\em Proof of Theorem} 1.1. 

  Related to the various definitions of conserved quantities for asymptotically 
AdS and asymptotically locally AdS spacetimes, there has been much recent 
discussion in the literature concerning the validity of the first law of 
black hole mechanics for such spacetimes, see in particular [34] and 
references therein and thereto. The ambiguities regarding the first law 
for the holographic mass (2.27) obtained by holographic or covariant 
renormalization have recently been resolved in a very clear analysis 
by Papadimitriou-Skenderis [19]. In particular, in the context of the present 
work, it is proved in [19] that if $g_{S} \in {\mathcal E}$ is globally 
stationary, then $g$ is a critical point for the holographic mass 
(2.27), under infinitesimal Einstein variations $h$ with fixed boundary 
metric, i.e. 
\begin{equation} \label{e3.5}
\delta m = 0, 
\end{equation}
among infinitesimal Einstein variations $h$ such that 
\begin{equation} \label{e3.6}
h_{(0)} = 0. 
\end{equation}
This is a special case of the first law of black hole mechanics in the 
AdS setting, namely in the case where there is no black hole, in the 
sense that the bifurcate Killing horizon is empty. 

\medskip

 We first observe that (3.5) also holds for all formal infinitesimal 
Einstein deformations $h_{f}$, i.e. formal series solutions of the linearized 
Einstein equations determined by $(h_{(0)}, \sigma)$, for $\sigma$ 
satisfying (2.32) on $(M, g_{S})$.  To see this, from (3.3) one has:
\begin{equation} \label{e3.7}
\int_{{\mathcal I}}\langle {\mathcal L}_{Z}\tau_{(n)}, h_{(0)} \rangle dV = 
-2\int_{{\mathcal I}}\langle \delta\sigma_{(n)}, X \rangle dV = 
-2[\delta m_{\partial\Sigma^{+}}(h) - \delta m_{\partial\Sigma^{-}}(h)]. 
\end{equation}
Here $h$ is any global linearized Einstein deformation with supp 
$h_{(0)}$ contained in the region of ${\mathcal I}$ between 
$\partial\Sigma^{-}$ and $\partial\Sigma^{+}$ and $\sigma_{(n)} = 
h_{(n)} - r_{(n)}'$. Since $h_{(0)} = 0$ on $\partial\Sigma^{-}$, by 
(3.5) one has
$$\delta m_{\partial\Sigma^{-}}(h) = 0. $$
Also, since $(M, g_{S})$ is stationary, the left side of (3.7) vanishes. 
On the other hand, (3.7) holds with $\sigma_{(n)}$ replaced by any 
$\sigma$ satisfying (2.32), so that 
\begin{equation} \label{e3.8}
\delta m_{\partial\Sigma^{+}}(h_{f}) = 0, 
\end{equation}
for any formal variation $h_{f}$ determined by $(h_{(0)}, \sigma)$. 

 Now suppose the solution $g \in {\mathcal E}$ is asymptotically stationary, 
i.e.~as $t_{i} \rightarrow \infty$, (or $t_{i} \rightarrow  -\infty$), 
$g_{t_{i}}$ converges to a stationary solution $g_{\infty}$. By the 
discussion above, on the limit $g_{S}$, one has
$$\delta m_{S}(h_{f}) = \int_{\partial\Sigma}\sigma (Z, \nu) = 0, $$
for any formal solution $h_{f} = (h_{(0)}, \sigma) \in {\mathcal F}$ to the 
linearized Einstein equations at ${\mathcal I}$ with $h_{(0)} = 0$ on 
$\partial \Sigma$, (or of compact support). It follows that on the original 
spacetime $(M, g)$, for $|T|$ sufficiently large, one has 
\begin{equation} \label{e3.9}
|\delta m_{T}(h)| = |\int_{\partial\Sigma_{T}}\sigma (Z, \nu)| \leq \varepsilon , 
\end{equation}
for all variations $(h_{(0)}, \sigma)$ which are uniformly bounded by 
a fixed constant, independent of $T$ and for which supp $h_{(0)}$ is a 
fixed compact set in ${\mathcal I}$. By Lemma 2.3, there exist such 
uniformly bounded $\sigma$, for all such variations $h_{(0)}$. 
Applying (3.4) once more to $(M, g)$ with the data $h_{f} = (h_{(0)}, \sigma)$ 
gives
\begin{equation} \label{e3.10}
\int_{{\mathcal I}}\langle {\mathcal L}_{Z}\tau_{(n)}, h_{(0)}\rangle dV = 
-2\int_{{\mathcal I}}\langle \delta\sigma , Z \rangle dV = 
-2[\delta m_{\partial\Sigma_{T^{+}}}(h_{f}) - 
\delta m_{\partial\Sigma_{T^{-}}}(h_{f})]. 
\end{equation}
Letting $T^{+} \rightarrow +\infty$ and $T^{-} \rightarrow -\infty$, 
using the boundedness condition, it follows from (3.9)-(3.10) that
\begin{equation} \label{e3.11}
\int_{{\mathcal I}}\langle {\mathcal L}_{Z}\tau_{(n)}, h_{(0)} \rangle dV = 0, 
\end{equation}
for all $h_{(0)}$ of compact support. Since $h_{(0)}$ is an arbitrary 
variation of the boundary metric $\gamma$ in ${\mathcal I}_{[0,T]}$, it 
follows that
\begin{equation} \label{e3.12}
{\mathcal L}_{Z}g_{(0)} = 0 \ {\rm and} \  {\mathcal L}_{Z}\tau_{(n)} = 0 
\ {\rm on} \  {\mathcal I} . 
\end{equation}
Thus, the Cauchy data $(g_{(0)}, \tau_{(n)})$ or $(g_{(0)}, g_{(n)})$ for 
$g$ on ${\mathcal I}$ are invariant under the flow of the Killing field $Z$ 
on ${\mathcal I}$. 

 We claim that (3.12) together with the unique continuation property at 
${\mathcal I}$ implies that $(M, g)$ is stationary. Intuitively, this is 
quite clear, but the proof requires some details. 

 To begin, extend $Z$ to a vector field in the bulk by the following 
two-step process. First, extend $Z$ to a neighhorhood $W$ of ${\mathcal I}$ 
in $M$ by requiring
$$[Z_{1}, \bar \nabla \rho ] = 0, $$
with $Z_{1}|_{{\mathcal I}} = Z$. Thus $Z$ is extended by the flow of 
$\bar \nabla\rho$ to a vector field $Z_{1}$, defined in the region where 
$\bar \nabla \rho$ is smooth. The corresponding form $\delta^{*}Z_{1}$ 
is then an infinitesimal Einstein deformation which preserves the 
defining function $\rho$. If $\phi_{s}$ denotes the flow of $Z_{1}$ 
and $g_{s} = \phi_{s}^{*}g$ denotes the corresponding curve of Einstein 
metrics, then the geodesic defining function for $g_{s}$ with fixed 
boundary metric $\gamma$ is the fixed function $\rho$. Each metric 
$g_{s}$ thus has the expansion (2.9)-(2.10) with fixed $\rho$. Since 
$\gamma$ is fixed, apriori only the $g_{(n),s}$ terms can vary, but 
(3.12) shows the variation of $g_{(n),s}$ vanishes to 1st order in $s$ at 
$s = 0$. Hence, since the formal series (2.9)-(2.10) are determined by the 
$(g_{(0)}, g_{(n)})$ terms, it follows that 
\begin{equation} \label{e3.13}
\delta^{*}Z_{1} = O(\rho^{p}), \ {\rm for \ all} \ p < \infty ,  
\end{equation}
formally near ${\mathcal I}$. More precisely, (3.13) holds to the extent that 
the metric $\bar g$ in (2.3) is smooth, (in a polyhomogeneous sense), up to 
${\mathcal I}$. In any case, one has
\begin{equation} \label{e3.14}
\delta^{*}Z_{1} = o(\rho^{n}). 
\end{equation}
Extend $Z_{1}$ outside $W$ arbitrarily but smoothly to all of $M$. 

 Next we need to bring $\delta^{*}Z_{1}$ into the transverse-gauge (2.22). 
To do this, (3.14) implies that 
$$\omega  \equiv \delta\bar \delta^{*}Z_{1} = o(\rho^{n}). $$
Set $Z = Z_{1}-Y$, where $Y$ is chosen to be the unique solution of the 
initial boundary value problem
\begin{equation} \label{e3.15}
\delta\bar \delta^{*}Y = \omega , 
\end{equation}
with zero Cauchy and boundary data; $Y|_{\Sigma} = 
\nabla_{\partial_{t}}Y|_{\Sigma} = 0$ and $Y|_{{\mathcal I}} = 0$. As 
discussed in \S 2, this equation has a unique smooth solution at least 
within the domain of dependence $D(\Sigma)$. The smallest indicial 
root of the operator $\delta\bar \delta^{*} = \frac{1}{2}D^{*}D - 
Ric$ is at least $(n+1)$, i.e.~the formal expansion of a solution of 
(3.15) has determined coefficients up to order $(n+1)$, cf.~[35]. Since 
$\omega = o(\rho^{n})$, it follows that 
$$\delta^{*}Y = o(\rho^{n}). $$
Hence the vector field $Z$ on $D(\Sigma)$ satisfies
\begin{equation} \label{e3.16}
\delta\bar \delta^{*}Z = 0, 
\end{equation}
with $Z|_{\partial \Sigma}$ the given Killing field $Z$ on ${\mathcal I}$ and 
\begin{equation} \label{e3.17}
\delta^{*}Z = o(\rho^{n}). 
\end{equation}

 Since $\delta^{*}Z$ is an infinitesimal Einstein deformation, it 
follows from (2.17) that $\delta^{*}Z$ satisfies the equations (2.22), 
i.e. 
\begin{equation} \label{e3.18}
D^{*}D(\delta^{*}Z) - 2R(\delta^{*}Z) = 0. 
\end{equation}
Since (3.17) holds, the unique continuation property at ${\mathcal I}$ for 
(3.18) implies that 
\begin{equation} \label{e3.19}
\delta^{*}Z = 0, 
\end{equation}
in a neighborhood $U$ of $\partial \Sigma$ in $D(\Sigma)$. 

 To show that $Z$ extends to a Killing field on all of $D(\Sigma)$, 
let $U_{s} = \{x\in D(\Sigma): \rho (x) \geq  s > 0\}$ and let $C_{s} 
= \partial U_{s}$, so that $C_{s}$ is a timelike cylinder. For $s$ 
sufficiently small, one has $\delta^{*}Z = 0$ to infinite order on 
$C_{s}$ and $\delta^{*}Z$ satisfies (3.18) throughout $D(\Sigma)$. The 
equations (3.18) are a hyperbolic system of PDEs, and at leading order 
are a diagonal system of scalar wave equations for which the boundary 
$C_{s}$ is strictly pseudoconvex. A unique continuation result of 
Tataru [36], then implies that $\delta^{*}Z = 0$ in a neighborhood of 
$C_{s}$ within $U_{s}$. One may then iterate this process a finite 
number of times to cover a neighborhood of the initial surface $\Sigma$. 
Of course one must change the distance function $\rho$ near regions 
where $\rho$ becomes singular and use instead smooth distance 
functions, but the arguments are otherwise the same. 

   Since $\delta^{*}Z$ thus vanishes to 1st order on all of $\Sigma$ 
and vanishes on ${\mathcal I}$, it follows from uniqueness of the initial 
boundary value problem for (3.18) that $\delta^{*}Z = 0$ on all of $M$. 

 Observe that $Z$ cannot become null anywhere in $M$. For if $Z$ is null 
at some point $p\in M$, then the Killing equation (3.19) implies that 
flow line $\sigma$ of $Z$ through $p$ is a null line, i.e. $Z$ remains 
null along $\sigma$. Since such null lines must intersect ${\mathcal I}$, 
this implies $Z$ is null somewhere on ${\mathcal I}$, which is impossible. 
Thus $Z$ is timelike throughout $M$, so that $(M, g)$ is globally 
stationary. 

{\endproof}

{\em Proof of Corollary} 1.2.

  Theorem 1.1 applies to any time-like Killing field on $({\mathcal I}, \gamma)$. 
Let ${\mathcal K}$ be the cone of time-like Killing fields on $({\mathcal I}, \gamma)$. 
Taking linear combinations, the cone ${\mathcal K}$ generates the full space of 
Killing fields on ${\mathcal I}$. Consider the space ${\mathcal S}_{\mathcal K}$ 
of solutions in ${\mathcal E}$ which are invariant under an effective 
${\mathcal K}$-action, restricting to the action of ${\mathcal K}$ at conformal 
infinity. Then Theorem 1.1 implies that any solution $(M, g) \in {\mathcal E}$ 
which to the future and past is weakly asymptotic to (possibly distinct) 
elements in ${\mathcal S}_{\mathcal K}$ is necessarily a fixed solution in 
${\mathcal S}_{\mathcal K}$. 

  Applying this to metrics in ${\mathcal E}$ whose conformal infinity is the 
Einstein static cylinder, it follows that the component of the identity of the 
isometry group of the Einstein static cylinder ${\mathbb R}\times S^{n-1}$ extends 
to a group of isometries of any $(M, g)\in {\mathcal E}$. In particular, any 
such $(M, g)$ has an isometric ${\mathbb R}\times SO(n)$ action. This implies 
that the Einstein equations (1.1) reduce to a system of ODE's and it is 
standard that the only globally smooth solution of this system is the exact AdS 
solution. 

{\endproof}

  Exactly the same arguments can be applied to spacetimes $(M, g)\in {\mathcal E}$ 
whose conformal infinity is homogeneous, i.e.~$({\mathcal I}, \gamma)$ has a 
transitive group of isometries. The Einstein equations for $(M, g)$ then 
reduce to a system of ODE's, (in the variable $\rho$), and the requirement 
that the solutions are smooth in the interior typically gives either a 
unique solution, or uniqueness up to a set of parameters which determine 
the topology of $(M, g)$. 

  To illustrate on some concrete examples, consider the AdS soliton metric 
of Horowitz-Myers [11]. In the toroidal compactification, $({\mathcal I}, \gamma)$ 
is the flat product metric on ${\mathbb R}\times T^{n-1}$ on the $(n-1)$-torus. 
This is clearly homogeneous. The corresponding ODE's may be solved explicitly 
and have a unique smooth solution on $(M, g) \simeq {\mathbb R}\times D^{2}\times T^{n-1} 
\in {\mathcal E}$ up to the choice of topology, (a choice of the disc $D^{2}$ bounding 
an $S^{1}\subset T^{n-1}$). This proves uniqueness of the AdS soliton metric 
among all (dynamical) metrics in ${\mathcal E}$ with the given conformal infinity 
and topology which are asymptotic at $t = +\infty$ and $t = -\infty$ to an AdS soliton 
metric. (This argument can be extended without difficulty to the case where 
${\mathcal I}$ is compactified to a single circle instead of the full $(n-1)$-torus). 
Exactly the same results hold for the recent AdS soliton metric analysed by 
Copsey-Horowitz [12]. 

  Similar uniqueness results also hold with respect to perturbations of such 
homogeneous conformal infinities. For example, suppose $(M, g)$ is an 
asymptotically simple, globally static solution of the Einstein equations 
(1.1) with conformal infinity $({\mathcal I}, \gamma)$ which is non-degenerate, 
(e.g.~$(M, g)$ has non-positive curvature). Given a static or stationary perturbation 
$({\mathcal I}, \widetilde \gamma)$ of the boundary data $({\mathcal I}, \gamma)$, 
there is a unique globally static or stationary asymptotically simple solution 
$(M, \widetilde g)$ close to $(M, g)$ with conformal infinity 
$({\mathcal I}, \widetilde \gamma)$, cf.~[37], [38]. Theorem 1.1 then 
implies the solution $(M, \widetilde g)$ is unique among all dynamical 
solutions to the Einstein equations in ${\mathcal E}$ which are asymptotic to the 
future and past to the given static or stationary solution. 

\begin{remark} \label{r3.1}

{\rm It is worth emphasizing that the results above require the solutions 
$(M, g)$ of the Einstein equations to be globally defined and non-singular. 
On the one hand, this is apparent from the proof. Theorem 1.1 uses the global 
vanishing (3.5) of the variation of mass on stationary spacetimes; per se, this 
is false if there are inner boundaries in addition to the boundary at conformal 
infinity. 

  In fact, Theorem 1.1 or Corollary 1.2 are false for solutions of the Einstein equations 
defined only in a neighborhood of conformal infinity $({\mathcal I}, \gamma)$. For 
example, let $({\mathcal I}, \gamma_{0})$ be the Einstein static cylinder and let 
$\tau_{(n)}$ be any analytic symmetric bilinear form on ${\mathcal I}$ which is 
asymptotic to 0 as $t \rightarrow \pm \infty$, and which satisfies the constraint 
equations (2.26). The result of Kichenassamy [22] mentioned above gives the existence 
of a solution of the Einstein equations (1.1) defined in a neighborhood of 
${\mathcal I}$ with the given $\tau_{(n)}$, i.e.~$g_{(n)}$ term, on ${\mathcal I}$. 
This solution is asymptotic to the exact AdS solution at $t = \pm \infty$, but 
is not exact AdS unless $\tau_{(n)} \equiv 0$. }
\end{remark}

\section{Discussion}
\setcounter{equation}{0}

  In the context of the Euclidean (or Riemannian) version of the AdS/CFT 
correspondence, it is important to know to what extent the boundary data 
$(\partial M, \gamma)$ determine the bulk solutions $(M, g)$ of the Einstein 
equations (1.1). Although it is possible in general that there are infinitely 
many topological types for $(M, g)$, or that, fixing the topology, the space 
of solutions has infinitely many components, there is only at most a finite 
dimensional moduli space of solutions when one fixes the topology and 
component. This follows essentially from the elliptic character of the 
Einstein equations (1.1). Thus, the ``Cauchy data'' $(g_{(0)}, g_{(n)})$ 
uniquely determine the bulk solution $(M, g)$ up to local isometry, (cf.~[10]), 
and given $g_{(0)}$, although the stress-energy term $g_{(n)}$ may not 
quite be uniquely determined by the boundary metric $g_{(0)}$, it is determined 
up to a finite dimensional space of parameters, (given a choice of topology and 
deformation component). 

  In this Euclidean context, the fact that isometries of the boundary 
$(\partial M, \gamma)$ necessarily extend to isometries of any smooth global 
bulk solution $(M, g)$ is a simple and clear illustration of (elementary or 
classical) aspects of the AdS/CFT correspondence. 

\medskip

  On the other hand, in the context of the Lorentzian solutions of the Einstein 
equations (1.1) in ${\mathcal E}$, there is an infinite dimensional space of normalizable 
modes, i.e.~$L^{2}$ solutions of the linearized Einstein equations. At least in some 
situations, for example that of exact AdS spacetime, these linearized solutions remain 
uniformly bounded in time, i.e.~$g_{AdS}$ is linearization stable. One expects that 
that $g_{AdS}$ is in fact dynamically stable, with the behavior of the nonlinear 
exact solutions nearby to $g_{AdS}$ well-modeled on the linearized behavior. 

  In fact, Friedrich [39] has shown that one may solve the 
initial boundary value problem for the Einstein equations (1.1) at least 
locally in time, in 3+1 dimensions. Thus, suppose the boundary data 
$({\mathcal I}, \gamma)$ are globally stationary, and that $\Sigma$ is a 
Cauchy surface with initial data $(g_{0}, K)$ satisfying the constraint 
equations (2.15) and matching $\gamma$ at the corner $\partial \Sigma 
\subset {\mathcal I}$, (cf.~also [40]). Then there is a solution $g$ of the 
Einstein equations (1.1), with conformal infinity $({\mathcal I}, \gamma)$ 
defined on a thickening $I\times \Sigma$ of $\Sigma$, realizing the given 
Cauchy data. This gives an infinite dimensional space of local-in-time 
exact solutions. If the data $(g_{0}, K)$ are (arbitrarily) close to the 
data $(g_{-1}, 0)$ of exact AdS spacetime, where $g_{-1}$ is the hyperbolic 
metric on the $n$-ball, then the resulting solution will exist for an 
(arbitrarily) long time, to the future and past. One would expect, although 
this remains to be proved, that such solutions extend to global-in-time 
solutions which remain globally asymptotically simple and globally close to 
$g_{AdS}$. 

  Theorem 1.1 implies that such global solutions cannot be both future 
and past asymptotic to the exact AdS spacetime. This is of course in strong 
contrast to the asymptotically flat situation $(\Lambda = 0)$, where the 
Christodoulou-Klainerman theorem [2] implies that small global perturbations 
of Minkowski spacetime disperse in time and tend to the flat solution, 
preserving, (at least to a certain degree), asymptotic simplicity and 
boundedness; in some situations asymptotic simplicity and boundedness is 
preserved to all orders, [5]. To a certain extent, it is the global conformal 
or causal structure which leads to these differences. Thus, observe that the 
Bondi mass on ${\mathcal I}^{+}$ is not preserved in time, but decreases 
monotonically. While the ADM mass is preserved under time evolution, it is 
defined with respect to the singular structure at spacelike infinity $\iota^{0}$, 
which does not match smoothly with the geometry of null infinity ${\mathcal I}^{+}$. 

  Similarly, with regard to the global stability results of de Sitter 
spacetimes in [8] and [9], here it is not clear if there is even a 
reasonable definition of mass, since ${\mathcal I}$ is spacelike. Given a 
definition of mass, as for instance in [41], it is not directly related 
to the dynamical evolution of the spacetime; it is a fixed quantity at 
future or past spatial infinity. 

  On the other hand, Theorem 1.1 is consistent with the known oscillatory behavior 
of the Einstein equations linearized at exact AdS, cf.~[1]. It is unknown if 
general asympotically locally AdS Einstein spacetimes are linearization 
stable, or what conditions will guarantee linearization stability. 

\medskip

  To conclude, we discuss informally the implication of Theorem 1.1 on the 
global dynamical behavior of bounded solutions in ${\mathcal E}$. Thus, consider 
the subspace ${\mathcal E}_{K}$ of uniformly bounded solutions in ${\mathcal E}$, 
i.e.~solutions which remain uniformly bounded by a fixed constant $K$ in time, 
in a suitable smooth norm. Then ${\mathcal E}_{K}$ is weakly compact, in that 
in any sequence in ${\mathcal E}_{K}$ has a subsequence converging in a weaker 
topology to a limit solution again in ${\mathcal E}_{K}$. Standard results in 
dynamical systems, (cf. [42] for example), then imply that the resulting Einstein 
flow on the constraint space ${\mathcal C}_{K}$ has (many) uniformly recurrent 
points: thus there exist times $t_{i} \rightarrow \infty$, (or $t_{i} \rightarrow 
-\infty$), with $t_{i+1}-t_{i} \leq T$, for some fixed $T$, and points 
$(g_{t_{i}}, K_{t_{i}}) \in {\mathcal C}_{K}$ such that 
$$dist((g_{t_{i+1}}, K_{t_{i+1}}), (g_{t_{i}}, K_{t_{i}})) \leq \varepsilon,$$
where the distance is taken in the weaker norm. Here $\varepsilon$ may be chosen 
arbitrarily small, but then with $T$ becoming possibly arbitrarily large. 

  Theorem 1.1 implies that no orbits of the flow on ${\mathcal C}_{K}$ are asymptotic 
to fixed points as $t \rightarrow \pm \infty$. This indicates that most all orbits 
in ${\mathcal C}_{K}$ should be almost periodic. 

  It is an open question whether there exist periodic (non-stationary) solutions in 
${\mathcal E}$. The proof of Theorem 1.1 can be easily adapted to rule out periodic 
solutions provided, for any periodic boundary variation $h_{(0)}$, there exists a 
periodic solution $\sigma$ of the linearized constraints as following (3.7). However, 
it is not clear that such periodic $\sigma$ always exist; I am very grateful to 
Bob Wald for discussions on this point.

\section*{Appendix}
\setcounter{equation}{0}
\begin{appendix}
\setcounter{section}{1}

  Let $X$ be a Killing field with respect to a Lorentz metric $({\mathcal I}, \gamma)$, 
and let $\tau$ be a divergence-free symmetric bilinear form on ${\mathcal I}$, 
$\delta \tau = 0$. We give here a proof of the identity (3.1) from [10], i.e.
\begin{equation}\label{eA.1}
\int_{{\mathcal I}}\langle {\mathcal L}_{X}\tau, \kappa \rangle dV = 
-2\int_{{\mathcal I}}\langle \delta'(\tau), X \rangle dV,
\end{equation}
where $\kappa = \frac{d}{ds}(\gamma + u\kappa)|_{s=0}$ is a variation of $\gamma$ of 
compact support and $\delta' = \frac{d}{ds}\delta_{\gamma + s\kappa}|_{s=0}$ is the 
variation of the divergence. The proof of (A.1) below holds for metrics of any 
signature. 

  To prove (A.1), we use the following standard formulas, cf.~[43] for example: 
\begin{equation} \label{eA.2}
{\mathcal L}_{V}\phi = \nabla_{V}\phi + 2\nabla V\circ \phi,
\end{equation}
\begin{equation}\label{eA.3}
(\delta^{*})'V = {\tfrac{1}{2}}\nabla_{V}\kappa + \delta^{*}V\circ \kappa,
\end{equation}
for any vector field $V$. Here $\phi \circ \psi$ is the symmetrized product; 
$(\phi \circ \psi)_{ij} = \frac{1}{2}(\langle \phi_{i}, \psi_{j} \rangle + 
\langle \phi_{j}, \psi_{i} \rangle)$ and $(\delta^{*}V)_{ij} = \frac{1}{2}
(V_{i,j} + V_{j,i})$. 

  To begin, by (A.2), $\int_{{\mathcal I}}\langle {\mathcal L}_{X}\tau, \kappa \rangle = 
\int_{{\mathcal I}}\langle \nabla_{X}\tau, \kappa \rangle + 
2\langle \nabla X \circ \tau, \kappa \rangle$. Since $\kappa$ is symmetric, 
$\langle \nabla X \circ \tau, \kappa \rangle = \langle \delta^{*}X 
\circ \tau, \kappa \rangle$. For the first term, write 
$\langle \nabla_{X}\tau, \kappa \rangle = X\langle \tau, \kappa \rangle 
- \langle \tau, \nabla_{X}\kappa \rangle$. The first term here integrates to 
$\delta X \langle \tau, \kappa \rangle$, (here we use the fact that $\kappa$ 
is of compact support), while by (A.3), the second term is 
$- \langle \tau, \nabla_{X}\kappa \rangle = -2 \langle \tau, 
(\delta^{*})'X \rangle + 2\langle \tau, \delta^{*}X\circ \kappa \rangle$. 
Hence 
$$\int_{{\mathcal I}}\langle {\mathcal L}_{X}\tau, \kappa \rangle dV = 
-2\int_{{\mathcal I}}\langle \tau, (\delta^{*})'X \rangle dV + 
4\int_{{\mathcal I}}\langle \tau,  \delta^{*}X \circ \kappa \rangle + 
\int_{{\mathcal I}}\delta X \langle \tau, \kappa \rangle dV .$$  

  Next, a straightforward computation using the fact that $\delta \tau = 0$ 
gives 
$$\int_{{\mathcal I}}\langle \tau, (\delta^{*})'X \rangle dV = 
\int_{{\mathcal I}}\langle (\delta')(\tau), X \rangle dV + 
2\int_{{\mathcal I}}[\langle \tau\circ \delta^{*}X,  \kappa \rangle - 
{\tfrac{1}{2}}\langle \tau,\delta^{*}X \rangle tr \kappa] dV :$$
the last two terms come from variation of the metric and volume form. 
Combining these computations gives 
\begin{equation}\label{eA.4}
\int_{{\mathcal I}}\langle {\mathcal L}_{X}\tau, \kappa \rangle dV = 
-2\int_{{\mathcal I}}\langle \delta'(\tau), X \rangle dV +  
\int_{{\mathcal I}}[\delta X \langle \tau, \kappa \rangle + 
\langle \tau, \delta^{*}X \rangle tr \kappa]dV .
\end{equation}
This gives (A.1) when $X$ is Killing, i.e. $\delta^{*}X = 0$. 

\medskip

  Let $\hat {\mathcal L}_{X}\gamma$ denote the conformal Killing operator: 
$\hat {\mathcal L}_{X}\gamma = {\mathcal L}_{X}\gamma - \frac{2div X}{n}\gamma$, 
(where $div X = -\delta X$). Then the calculations above give, (again for 
divergence-free $\tau$), 
\begin{equation}\label{eA.5}
\int_{{\mathcal I}}\langle {\mathcal L}_{X}\tau + div X \tau, \kappa \rangle dV = 
-2\int_{{\mathcal I}}\langle \delta'(\tau), X \rangle dV + {\tfrac{1}{2}}
\int_{\mathcal I}\langle \tau, \hat {\mathcal L}_{X}\gamma \rangle dV + 
{\tfrac{1}{n}}\int_{\mathcal I} div X \,a \cdot tr \, \kappa dV, 
\end{equation}
where $a = tr \tau$. If $\sigma$ satisfies the linearized constraint equations (2.32) 
then a simple calculation from (A.5), gives
\begin{equation}\label{eA.6}
\int_{{\mathcal I}}\langle {\mathcal L}_{X}\tau + [1-{\tfrac{2}{n}}]div X \,\tau, \kappa 
\rangle dV = 
\end{equation}
$$\int_{\mathcal I}\langle \sigma + {\tfrac{1}{2}}tr \kappa \, \tau, 
\hat {\mathcal L}_{X}\gamma \rangle dV - 2\int_{\mathcal I}div (\sigma(X)) dV + 
{\tfrac{1}{n}}\int_{\mathcal I} div X (a\cdot tr \, \kappa + 2a') dV.$$ 
When $X$ is a conformal Killing field on $({\mathcal I}, \gamma)$, the first term 
on the right vanishes. When $\tau = \tau_{(n)}$, the second term gives 
the difference of the variation of the mass when $X$ is timelike by the 
divergence theorem while the integrand on the left is related to the 
transformation properties of the stress-energy tensor $\tau_{(n)}$ under conformal 
changes, cf.~[17].

\end{appendix}

\bibliographystyle{plain}

\begin{thebibliography}{WWW}

\footnotesize

\bibitem [1]{1} A. Ishibashi and R. Wald, Dynamics in non-globally-hyperbolic static 
spacetimes: III. Anti-de Sitter spacetime, Class. Quant. Grav. {\bf 21}, (2004), 
2981-3014, [arXiv:hep-th/0402184]. 

\bibitem [2]{2} D. Christodoulou and S. Klainerman, The Global Nonlinear Stabilty 
of Minkowski Space, Princeton Univ. Press, Princeton, 1993. 

\bibitem [3]{3} S. Klainerman and F. Nicolo, The Evolution Problem in General Relativity, 
Birkh\"auser Verlag, Boston, (2003). 

\bibitem [4]{4} H. Lindblad and I. Rodnianski, The global stability of the Minkowski 
space-time in harmonic gauge, Annals of Math, (to appear), [arXiv:math.AP/0411109]. 

\bibitem [5]{5} P. Chru\'sciel and E. Delay, Existence of non-trivial, vacuum 
asymptotically simple space-times, Class. Quant. Grav. {\bf 19}, (2002), L71-L79, 
[arXiv:gr-qc/0203053]. 

\bibitem [6]{6} L. Andersson and V. Moncrief, Future complete vacuum spacetimes, in: 
The Einstein Equations and the Large Scale Behavior of Gravitational Fields, Ed: P. 
Chru\'sciel and H. Friedrich, Birkh\"auser Verlag, (2004), 299-330, 
[arXiv:gr-qc/0303045]. 

\bibitem [7]{7} Y. Choquet-Bruhat and V. Moncrief, Future global in time  Einsteinian 
spacetimes with $U(1)$ isometry group, Ann. H. Poincar\'e, {\bf 2}, (2001), 1007-1064, 
[arXiv:gr-qc/0112049]. 

\bibitem [8]{8} H. Friedrich, On the existence of $n$-geodesically complete or future 
complete solutions of Einstein's field equations with smooth asymptotic structure, 
Comm. Math. Phys. {\bf 107}, (1986), 587-609. 

\bibitem [9]{9} M. Anderson, Existence and stabilty of even dimensional asymptotically 
de Sitter spaces, Annals H. Poincar\'e, {\bf 6}, (2005), 801-820, [arXiv:gr-qc/0408072]. 

\bibitem [10]{10} M. Anderson, Unique continuation results for Ricci curvature and applications, 
[arXiv:math.DG/0501067]. 

\bibitem [11]{11} G. Horowitz and R. Myers, The AdS/CFT correspondence and a new positive 
energy conjecture for general relativity, Phys. Rev. {\bf D59}, (1999), 026005, 
[arXiv: hep-th/9808079]. 

\bibitem [12]{12} G. Horowitz and K. Copsey, Gravity dual of gauge theory on 
$S^{2}\times S^{1}\times {\mathbb R}$, [arXiv:hep-th/0602003]

\bibitem [13]{13} G. Galloway, S. Surya and E. Woolgar, On the geometry and mass of 
static, asymptotically AdS spacetimes, and the uniqueness of the AdS soliton, 
Comm. Math. Phys. {\bf 241}, (2003), 1-25, [arXiv:hep-th/0204081]. 

\bibitem [14]{14} M. Anderson, Geometric aspects of the AdS/CFT correspondence, 
in: AdS/CFT Correspondence: Einstein Metrics and their Conformal Boundaries, Ed. 
O. Biquard, (2005), 1-31, Euro. Math. Soc, Z\"urich, [arXiv:hep-th/0403087]. 

\bibitem [15]{15} E. Witten Anti de Sitter space and holography, 
Adv. Theor. Math. Phys. {\bf 2}, (1998), 253-291, [arXiv:hep-th/9802150]. 

\bibitem [16]{16} V. Balasubramanian and P. Kraus, A stress tensor for anti-de Sitter 
gravity, Comm. Math. Phys. {\bf 208}, (1999), 413-428  [arXiv:hep-th/9902121]

\bibitem [17]{17} S. de Haro, S. Solodukhin and K. Skenderis, Holographic reconstruction 
of spacetime and renormalization in the AdS/CFT correspondence, Comm. Math. Phys. 
{\bf 217}, (2001), 595-622, [arXiv:hep-th/0002230]. 

\bibitem [18]{18} I. Papadimitriou and K. Skenderis, AdS/CFT correspondence and geometry, 
in: AdS/CFT Correspondence: Einstein Metrics and their Conformal Boundaries, Ed. 
O. Biquard, (2005), 73-101, Euro. Math. Soc, Z\"urich, [arXiv:hep-th/0404176].

\bibitem [19]{19} I. Papadimitriou and K. Skenderis, Thermodynamics of asymptotically 
locally AdS spacetimes, [arXiv: hep-th/0505190]. 

\bibitem [20]{20} C. Fefferman and C.R. Graham, Conformal invariants, in: 
\'Elie Cartan et les Math\'ematiques d'Aujourd'hui, Asterisque, (1985), (Num\'ero hors 
s\'erie), 95-116 

\bibitem [21]{21} C. Imbimbo, A. Schwimmer, S. Theisen and S. Yankielowicz, 
Class. Quant. Grav. {\bf 17}, (2000), 1129, [arXiv:hep-th/9910267]. 

\bibitem [22]{22} S. Kichenassamy, On a conjecture of Fefferman and Graham, 
Adv. Math. {\bf 185}, (2004), 268-288. 

\bibitem [23]{23} R. Wald, General Relativity, Univ. Chicago Press, Chicago, (1984). 

\bibitem [24]{24} S. Hawking and G. Ellis, The Large Scale Structure of Space-Time, 
Cambridge Univ. Press, Cambridge, (1973). 

\bibitem [25]{25} L. Andersson and P. Chru\'sciel, On asymptotic behavior of solutions 
of the constraint equations in general relativity with ``hyperboloidal boundary 
conditions'', Dissertationes Math. {\bf 355}, (1996), 1-100. 

\bibitem [26]{26} M. Henningson and K. Skenderis, The holographic Weyl anomaly, 
JHEP, {\bf 9807}, 023, (1998), [arXiv:hep-th/9806087], Holography and the 
Weyl anomaly, Fortsch. Phys, {\bf 48}, (2000), 125, [arXiv:hep-th/9812032]. 

\bibitem [27]{27} J. deBoer, E. Verlinde and H. Verlinde, On the holographic 
renormalization group, JHEP {\bf 0008}, (2000), 003, [arXIv:hep-th/9912012]. 

\bibitem [28]{28} D. Martelli and W. M\"uck, Holographic renormalization and 
Ward identities with the Hamilton-Jacobi method, Nucl. Phys. {\bf B654}, (2003), 
248-276, [arXiv: hep-th/0205061]. 

\bibitem [29]{29} L. Abbott and S. Deser, Stability of gravity with a 
cosmological constant, Nucl. Phys. {\bf B195}, (1982), 76 

\bibitem [30]{30} A. Ashtekar and A. Magnon, Asymptotically anti-de Sitter 
spacetimes, Class. Quantum Gravity, {\bf 1}, (1984), L39-L44 

\bibitem [31]{31} M. Henneaux and C. Teitelboim, Aymptotically anti-de Sitter 
spaces, Comm. Math. Phys. {\bf 98}, (1985), 391

\bibitem [32]{32} S. Hollands, A. Ishibashi and D. Marolf, Comparison between 
various notions of conserved charges in asymptotically AdS-spacetimes, 
Class. Quant. Grav. {\bf 22}, (2005), 2881-2920, [arXiv:hep-th/0503045]. 

\bibitem [33]{33} S. Hollands, A. Ishibashi and D. Marolf, Counter-term 
charges generate bulk symmetries, Phys.Rev.{\bf D72}, (2005), 104025, 
[arXiv:hep-th/0503105].

\bibitem [34]{34} G. Gibbons, M. Perry and C. Pope, The first law of 
thermodynamics for Kerr-anti-de Sitter black holes, Class. Quant. Grav. {\bf 22}, 
(2005), 1503-1526, [arXiv:hep-th/0408217]. 

\bibitem [35]{35} J. Lee, Fredholm operators and Einstein metrics on conformally 
compact manifolds, [arXiv:math.DG/0105046]. 

\bibitem [36]{36} D. Tataru, Carleman estimates, unique continuation and applications, 
(preprint, 1999), http://www.math.berkeley.edu/ $\sim$ tataru/ucp.html

\bibitem [37]{37} M. Anderson, P. Chru\'sciel and E. Delay, 
Non-trivial static, geodesically complete vacuum space-times with a negative 
cosmological constant, JHEP, {\bf 10}, (2002), 063, [arXiv:gr-qc/0211006]

\bibitem [38]{38} P. Chru\'sciel and E. Delay, Non-singular, vacuum, stationary 
space-times with a negative cosmological constant, [arXiv:gr-qc/0512110]. 

\bibitem [39]{39} H. Friedrich, Einstein equations and conformal structure: 
Existence of anti-de Sitter space-times, Jour. Geom. Phys. {\bf 17}, (1995), 
125-184. 

\bibitem [40]{40} J. Kannar, Hyperboloidal initial data for the vacuum Einstein 
equations with cosmological constant, Class. Quant. Grav. {\bf 13}, (1996), 
3075-3084. 

\bibitem [41]{41} V. Balasubramanian, J. de Boer, and D. Minic, Mass, entropy 
and holography in asymptotically de Sitter spaces, Phys. Rev. {\bf D65}, (2002), 
123508, [arXiv:hep-th/0110108] 

\bibitem [42]{42} H. Furstenburg, Recurrence in Ergodic Theory and Combinatorial 
Number Theory, M.B. Porter Lectures, Princeton Univ. Press, Princeton, N.J. (1981). 

\bibitem [43]{43} A. Besse, Einstein Manifolds, Springer Verlag, 
New York, (1987). 


\end{thebibliography}

\bigskip

\begin{center}
August, 2006
\end{center}

\medskip
\noindent
{\address Department of Mathematics\\
S.U.N.Y. at Stony Brook\\
Stony Brook, NY 11790-3651}

\noindent
{Email: anderson@math.sunysb.edu}

\end{document}